\documentclass[aps,twocolumn,amsmath,amssymb]{revtex4}

\usepackage{graphicx}
\usepackage{dcolumn}
\usepackage{bm}

\usepackage{epstopdf}
\usepackage{longtable}

\begin{document}

\title{Fractional Quantum Hall Effect in the Second Landau Level}

\author{H.C. Choi$^{1}$, W. Kang$^{1}$, S. Das Sarma$^{2}$, L.N. Pfeiffer$^{3}$, and  K.W. West$^{3}$}

\affiliation{$^{1}$James Franck Institute and Department of Physics,  University of Chicago, Chicago, Illinois 60637\\
$^{2}$ Condensed Matter Theory Center, Department of Physics, University of Maryland, College Park, MD 20742\\
 $^{3}$Bell Laboratories, Alcatel-Lucent, 600  Mountain Avenue, Murray Hill, NJ 07974}


\begin{abstract}
We present activation gap measurements of the fractional quantum Hall
effect (FQHE) in the second Landau level. Signatures for 14
(5) distinct incompressible FQHE states are seen in a high (low)
mobility sample with the enigmatic 5/2 even denominator FQHE having
a large activation gap of $\sim$600 ($\sim$300mK) in the high (low)
mobility sample. Our measured large relative gaps for 5/2, 7/3, and
8/3 FQHE indicate emergence of exotic  FQHE correlations in the
second Ladau level, possibly quite different from the well-known lowest
Landau level Laughlin correlations. Our measured 5/2 gap is found to
be in reasonable agreement with the theoretical gap once finite
width and disorder broadening corrections are taken into account.
\end{abstract}

\maketitle

The clean (i.e. high mobility) two-dimensional electron system
(2DES) at low temperatures and high magnetic fields exhibits  a rich
array of exotic, highly correlated incompressible ground states. In
the lowest Landau level (LLL), the physics is dominated by the
sequence of FQHE states at odd denominator filling fractions with
more than 50 FQHE states with odd denominators  as large as 19
observed so far. The Laughlin wave function describes  the primary
FQHE states at LLL filling fractions  $\nu = 1/m, m = 3, 5, 7...$ as
an incompressible quantum fluid of electrons\cite{Laughlin83}. The
Laughlin states feature a gap in the energy spectrum with
fractionally charged quasiparticles with charge $q = \pm e/m$ as the
lowest energy excitation. The sequence of hierarchical $\nu =
p/(2p \pm 1), p = 1, 2, 3...$ higher order FQHE states is described
by the composite fermion model\cite{Jain89,Halperin93}. The odd
denominator constraint arises from
the antisymmetry of the many-body wave functions required under the
exchange of two electrons\cite{Laughlin83}. To date no even
denominator FQHE state has been observed in the LLL for a
single-layered 2DES, although certain anomalies have been observed
at $\nu = 3/8$ \cite{Pan03}.

The startling exception to the odd denominator rule is the
even denominator FQHE at $\nu = 5/2 = (2+ 1/2)$ in the second Landau
level(SLL). Early experiments\cite{Willett87} showed a weakly formed
quantized Hall plateau with a finite longitudinal resistance at low
temperatures. Improvement in the sample quality led to the formation
of a fully quantized Hall plateau along with a vanishing
longitudinal resistance at low
temperatures\cite{Pan99,Xia04,Eisenstein02}. 
In contrast to the LLL,
the SLL features an array of competing ground states including odd
denominator FQHE, reentrant insulating states, and even denominator
FQHE at $\nu$ = 5/2 and 19/8 = (2+ 3/8)\cite{Xia04}.

The theoretical understanding of the even-denominator FQHE at $\nu =
5/2$ is based on the p-wave pairing of composite fermions, similar
to the pairing in a chiral $p$-wave BCS
superconductor\cite{Moore91,Greiter91}. The variational wave
function for the paired Hall states is  modified by a Pfaffian that
creates a FQHE state. Numerical diagonalization calculations provide
strong support for the Pfaffian state as the ground state at  $\nu =
5/2$\cite{Morf98,Rezayi00}. The non-Abelian quasiparticle statistics
of the Pfaffian 5/2 state has received much attention recently for
the prospect of realizing topologically protected
qubits\cite{DasSarma05}. The existence of the even denominator FQHE
state and the general paucity of a large number of odd denominator
fractions clearly differentiate the FQHE physics of the SLL from
that in the LLL. In particular, as we demonstrate in this paper, the
standard composite fermion LLL hierarchy states seem to be strongly
suppressed in the SLL. The nature of SLL interaction and correlation
are not well-understood and the 5/2 state, although it is an even
denominator state with no analogy in the LLL, is both the best
understood and the strongest FQHE state in the SLL. In fact,
theoretical work\cite{SLLtheory} indicates that only $\nu <$ 2 + 1/3
Laughlin states would be stable in the SLL!

In this paper, we report on the observation of a large ($\sim$14)
number of possible incompressible states and their energy
gaps in the second Landau level. We find that the energy gap of the
$\nu$ = 5/2 FQHE states exceeds 500 mK in the high mobility samples.
Comparing results from two samples with "high" and "low" mobilities,
we conclude that in general the $\nu$ = 5/2 state is the strongest
FQHE in the SLL, followed by the 7/3 and 8/3 states 
with all other fractions being far weaker. The fact that an even
denominator fraction, considered to be a p-wave paired  Hall state, is the
strongest FQHE state in the SLL provides a sharp contrast between
the FQHE physics in the LLL and the SLL. We find the 7/3 and 8/3 states to be 
much stronger than the other odd denominator FQHE states in the SLL.

Two  symmetrically $\delta$-doped 30 nm wide quantum well samples with identical structures were studied.
The mobility for sample A (high-mobility) is $\mu = 28.3\times
10^{6}$cm$^{2}$/Vs with the electron density of $n = 3.2 \times
10^{11}$cm$^{-2}$. The mobility for sample B  (low-mobility) is  $\mu
= 10.5\times 10^{6}$cm$^{2}$/Vs with the electron density of $n =
2.8 \times 10^{11}$cm$^{-2}$.  The samples in a van der Pauw geometry were attached to
the cold finger of a dilution refrigerator. Magnetotransport studies
were made after illuminating the specimens with a red light emitting
diode at 4K.  Measurement was made using a low-frequency AC lock-in
technique at low temperatures down to $\sim$30mK.

\begin{figure}
\includegraphics[width=3.2in]{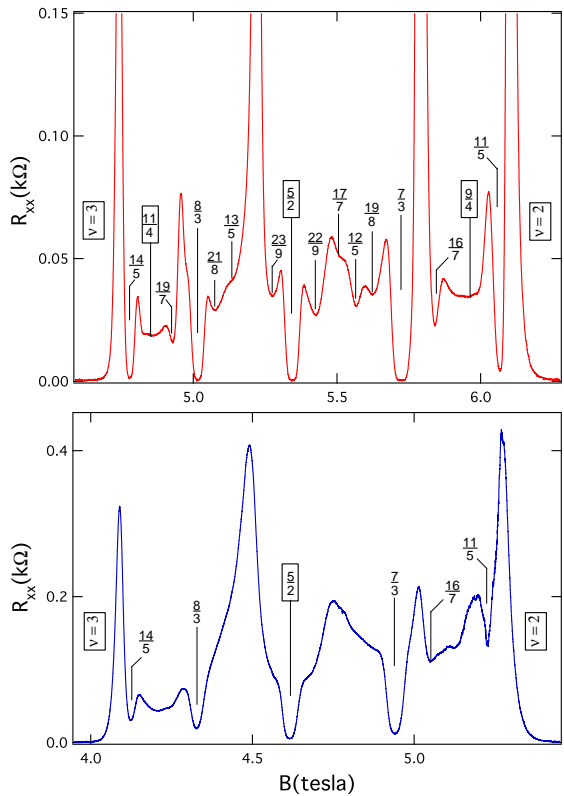}
\caption{Magnetoresistance in the second Landau level of a two-dimensional electron systems with (a) the mobility of $\mu = 28\times10^{6}$cm$^{2}$/Vs (sample A ) and (b) mobility of $\mu = 10.5\times10^{6}$cm$^{2}$/Vs. The temperature was 36 mK for both samples. \label{fig:Rxxhighlow}}
\end{figure}

Figs. \ref{fig:Rxxhighlow}a and  \ref{fig:Rxxhighlow}b show the low
temperature magnetoresistance for the sample A  and sample B. In the
higher mobility sample A, a remarkable array of 14 different FQHE
states are observed, as reflected in well-defined  $\rho_{xx}$
minima even at a relatively moderate temperature of 36 mK. The
most prominent FQHE states are found at fillings $\nu$ = 5/2, 7/3,
8/3, 14/5, 11/5, 12/5, 16/7, and 19/7. Additional  features may
be attributed to $\nu$ = 13/5, 17/7, 22/9, and 23/9 states.
 Finally, resistance minima can
be seen at even denominator fillings $\nu$ = 19/8 and 21/8. In
addition to these FQHE states, signatures of  reentrant insulating
states of varying strengths can be detected at $\nu$ = 2.69, 2.58, 2.44, and 2.31.
The large resistance arising from the reentrant insulating states weakens
the FQHE states found at $\nu$ = 13/5 and 17/7.

Fig. \ref{fig:Rxxhighlow}b shows that only five of the 14 states
seen in sample A survive in sample B: the FQHE states at $\nu$ =
5/2, 7/3, 8/3, 11/5, and 14/5. This is presumably a direct
suppression due to disorder since the two samples are very
similar except for a factor of $\sim$3 difference in mobility. Along
with the absence of many higher order FQHE states, the magnetoresistance
minima of the principal FQHE states have become much weaker in
sample B compared with sample A. A cursory look at our Fig.
\ref{fig:Rxxhighlow} immediately makes it obvious that the strongest
SLL FQHE states occur at $\nu$ = 5/2, 7/3, and 8/3 with the 5/2 and the
7/3 states being comparable, and the 8/3 state being somewhat
weaker. 

 \begin{figure}
\includegraphics[width=3.2in]{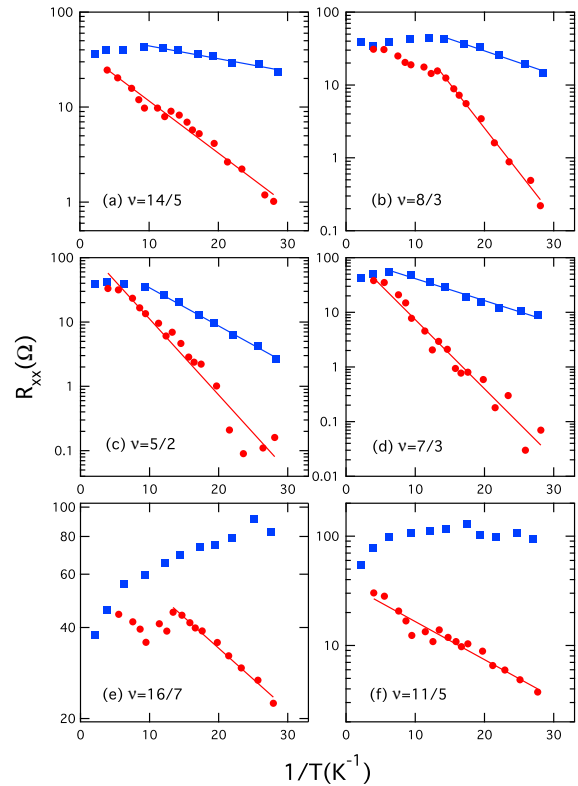}
\caption{Temperature dependence of R$_{xx}$ for various fractions in the second Landau level for samples A (solid circle) and B (solid square).
\label{fig:RxxTdep}}
\end{figure}

Figs. \ref{fig:RxxTdep}a-\ref{fig:RxxTdep}f show the temperature
dependence of the magnetoresistance mimina at filling fractions
$\nu$ = 14/5, 8/3, 5/2, 7/3, 16/7, and 11/5, respectively, for
samples A and B. Except for the highest temperatures where activated
behavior is not expected, the Arrhenius plot shows that the regions
of activation extend well over one decade in sample A  for $\nu$ =
14/5, 8/3, 5/2, and 7/3. Other FQHE states at $\nu$ = 16/7,
and 11/5 in sample A  show a more limited activation over the
measured temperature range. The FQHE states in sample B  show a
demonstrably weaker activation at same fillings compared with sample A.

\begin{table*}
\caption{\label{tab:energygap}
Energy gap measured for the fractional quantum Hall states in the second Landau level.}
\begin{ruledtabular}
\begin{tabular}{ccccccccc}
  &     &  $\nu  = 14/5$ &  $\nu  = 19/7$   & $\nu  =  8/3$ & $\nu  =  5/2$  &  $\nu  =  7/3$ & $\nu  =  16/7$ & $\nu  = 11/5$  \\
\hline
 sample A   & $\Delta$(mK) & 252  &  108   & 562  & 544  & 584 &  94  & 160 \\
 & $\Delta \left(\frac{e^{2}}{\epsilon \ell}\right)$ & 0.0023 &  0.0010 & 0.0050 &  0.0047 & 0.0049 &  0.0008 & 0.0013 \\
sample B  & $\Delta$(mK) & $\le60$ &    & 150 & 272  & 206 &  & $\le 40$ \\
 & $\Delta \left(\frac{e^{2}}{\epsilon \ell}\right)$  &    &  & 0.0014  &  0.0026  & 0.0019 &  &  \\
\end{tabular}
\end{ruledtabular}
\end{table*}

The energy gap $\Delta$ for the various FQHE states can be
determined from the Arrhenius plot using the activated resistance
R$_{xx}\propto exp(-\Delta/2T)$. While a greater range of activation
behavior is desired for sample B  and the higher order fractions in
sample A, the data for sample A with strong activation behavior
allows us to constraint the energy gap values for the states that
exhibit limited activation. For sample A at $\nu = 8/3$, an
anomalous change in the slope was observed near $\sim$ 80mK. The
origin of this puzzling feature is unclear.
No energy gaps were obtained for sample B  at $\nu$ = 16/7 and 11/5.
Results of our activation analysis are summarized in Table \ref{tab:energygap}. 
The energy gaps for $\nu$ = 8/3, 5/2, and 7/3 in sample A  are the
largest with their magnitudes exceeding 500mK. 
In sample B the 5/2 FQHE state is by far the
strongest incompressible state with the largest energy gap.

\begin{figure}
\includegraphics[width=3.2in]{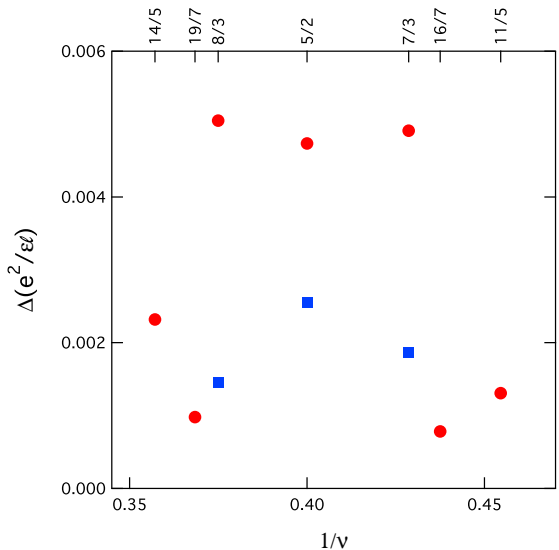}
\caption{Energy gaps for the fractional quantum Hall effect states in the second Landau level in the units of Coulomb energy $e^{2}/\epsilon\ell$, where $\epsilon$ is the dielectric constant  and $\ell = \sqrt{\hbar/eB}$ is  the magnetic length. Solid dots (squares) represent the energy gaps from sample A  (B). \label{fig:Gaps}}
\end{figure}

Fig. \ref{fig:Gaps} shows the energy gaps  at different fillings for the two samples. 
The gap magnitudes have been
converted to the units of Coulomb energy $e^{2}/\epsilon\ell$, where
$\epsilon$ is the background dielectric constant  and  $\ell =
\sqrt{\hbar/eB}$ is the magnetic length. 
The gaps for the $\nu$ = 8/3, 5/2, and 7/3 states approach
$\sim 0.005e^{2}/\epsilon\ell$, which is roughly an order of
magnitude smaller  than the corresponding gap values
for the strongest FQHE states ($\nu$ = 1/3, 2/3) in the LLL. The
energy gaps measured for sample B, compared with sample A, show 
that there is a strong suppression of the gaps with increasing
disorder. In the Coulombic energy scale, the gaps for the
$\nu$ = 5/2, 7/3, and 8/3 states are respectively reduced by approximately
50\%, 60\%, and 70\% between samples A and B. We note that disorder 
apparently affects the odd denominator states more strongly than the $\nu = 5/2$ state.

One of the more intriguing features of the data is that energy gap 
of  the $\nu$ = 7/3 and 8/3 states are disproportionally larger than
what may be expected under the standard model of FQHE in the LLL.  
Within the composite fermion model, the FQHE states at  $\nu$ = 2/5 and 
3/5 are the strongest after $\nu$ = 1/3 and 2/3 FQHE states.
The LLL energy gaps for the $\nu$ = 2/5 and 3/5 states
are approximately half of the gaps for the $\nu = 1/3$ and 2/3 states.
The gaps for the higher order FQHE states decrease proportionally with
their position in the sequence \cite{Du93}. 
Based on the relative strengths of the $\nu$ = 7/3 and 8/3 states compared 
to the $\nu$ = 12/5 and 13/5 states, the gaps for the $\nu$ = 7/3 
amd 8/3  states are approximately 8-10 times larger than the $\nu$ = 12/5 and
13/5 states.  This means that measured energy gaps for the $\nu$ = 7/3 and 
8/3 states are anomalously enhanced compared to to what may be expected under 
the composite fermion model. The enhanced gaps of the 7/3 and 8/3 states may be
related to their anomalous angular dependence \cite{Csathy05}.

Our finding that the strongest SLL incompressible state are 5/2,
7/3, 8/3, 14/5, 11/5 should be contrasted with the strongest LLL
incompressible states: 1/3, 2/3, 2/5, 3/5, 2/7. Apart from the
enigmatic 5/2 even denominator SLL
state, which has no analog in the LLL, there are several additional
features of the SLL FQHE states worth
emphasizing: (1) the  weakness of heirarchical
(e.g. 2+2/5 2+3/5) states; (2) the dominance
of primary fractions, e.g. 2+1/3, 2+1/5, 2+2/3, 2+4/5; (3) the relative
strength, $\Delta^{7/3, 8/3}/\Delta^{11/5, 14/5}\sim$ 2-3, of the
1/3 SLL state compared with the 1/5 SLL state. All three of these
SLL features are in sharp contrast with the LLL, where the
hierarchy states (i.e. 2/5, 3/5, etc.) are the strongest fractions
after the 1/3 state and the observed $\Delta_{1/3}/\Delta_{1/5}\le$
10 \cite{Jiang90} whereas the observed
$\Delta_{1/3}/\Delta_{2/5}\sim$ 2 \cite{Du93}. We find
$\Delta_{7/3}/\Delta_{11/5}\sim$ 2 whereas
$\Delta_{7/3}/\Delta_{12/5}\sim$ 10 using the activation gap of 70
mK quoted in ref. \onlinecite{Xia04} for the 12/5 state. Thus our observed 7/3,
8/3 states are "too strong" compared with the 12/5, 13/5 states,
but are "too weak" compared with the 11/5, 14/5 states, exactly
the reverse of the situation in the LLL.

Based on the above observations, we conclude that the 7/3, 8/3
states are unlikely to be the SLL analogs of the 1/3, 2/3 LLL
Laughlin correlated states whereas our observed 11/5, 14/5 states
are likely to be Laughlin states. This conclusion is consistent with
several theoretical predictions\cite{SLLtheory} 
where the Laughlin states are found to
be the unstable for Coulomb interaction in the SLL for $\nu$ =2+1/3.
We therefore believe it to be likely that all three strongly
incompressible SLL states (i.e. 5/2, 7/3, 8/3) are exotic
non-Laughlin FQHE states. This conclusion is also consistent with the proposal of
the weak 12/5, 13/5 SLL states being
parafermionic Read-Rezayi states\cite{Read99} rather than the garden
variety hierarchy states. It seems, therefore, that the SLL 
correlations are much more subtle than the LLL  correlations.

The best current theoretical estimate for the infinite system
extrapolated excitation gap for the 5/2 incompressible state is
$\Delta_{ex} \approx$ 0.025 in the  
Coulomb energy unit\cite{Morf, Feigun}. This ideal gap value requires (at least)
three corrections due to the finite width\cite{Morf} of the quasi-2D
system, disorder\cite{Morf}, and Landau level coupling\cite{Wojs06}, 
before any comparison with experiment can be made.
All three corrections suppress the theoretical gap
with the suppression due to the finite width effect, which softens
the Coulomb interaction, being the easiest to calculate.  The finite
width correction depends\cite{ Morf} on the parameter $w/\ell$
where $w$ is the quantum well width ($w$ =30 nm for both samples A
and B). Using the applied magnetic field values (5.3T for A and 4.6T
for B) we find $w/\ell\approx$ 2.7 (sample A) and 2.50 (sample B).
Such large values of $w/\ell$ imply rather strong finite width
corrections\cite{Morf} reducing$\Delta_{ex}$ at $\nu$ = 5/2 by a
factor of 2 or more  to about 0.013, 
which corresponds to a gap of 1.5K (sample A) and 1.4K (sample B).
Our observed 5/2 activation gaps $\Delta$ = 0.54K (sample A); 0.27K (sample
B) are substantially below the ideal gap values
because of disorder (and, possibly, Landau level
mixing), effects which are difficult to treat theoretically.  We
ignore Landau level coupling effects, although it may very well not
be negligible in reality, based on the argument that
$(e^{2}/\epsilon\ell_{1})/\hbar\omega_{c} \approx$ 0.4 is small,
where $\ell_{1}= \sqrt{(2n+1)}\ell$ is the Landau radius in the
$n$=1 Landau level.

The inclusion of disorder in the theory of FQHE gap is problematic
in the absence of a true transport theory. A simple procedure, used
extensively if somewhat unjustifiably, is to write the
disorder-induced gap as $\Delta \equiv \Delta_{ex}-\Gamma$ where
$\Gamma$ is the calculated level broadening. We can
theoretically estimate the zero-field level broadening of samples A
and B by using the sample structures ($w$ = 30nm with a spacer layer
of $d$ = 80nm) and the mobilities to get $\Gamma_{A} \approx$ 0.38K,
$\Gamma_{B} \approx$ 0.63K, where the level broadening $\Gamma$
corresponds to the so-called quantum single-particle impurity
broadening ($\Gamma_{s} \equiv \hbar/2\tau_{s}$) rather than the
transport mobility broadening ($\Gamma_{t} \equiv \hbar/2\tau_{t}$).
It is well-known\cite{dS85} that in high-mobility modulation-doped
structures $\tau_{t}/\tau_{s} \gg 1$, and in fact, for the
high-mobility structures used in our experiments $\Gamma_{s} \approx
200\Gamma_{t}$ due to the very large values of
$q_{s}d \sim 20$, where $q_{s}$ is  the screening wave vector.

Incorporating disorder (and finite width) correction in the
theoretical gap values we arrive at the following predictions for
the activation gaps: $\Delta_{A}$ = 1.5K - 0.8K $\approx$ 0.7K;
$\Delta_{B}$ = 1.4K - 1.2K $\approx$ 0.2K, which are comparable with our
experimentally measured gaps of 0.54K and 0.27K, respectively. We
note that much  of the suppression (a factor of
2) of the measured 5/2 gap ($\sim$ 0.005) compared with the
theoretical 5/2 gap ($\sim$ 0.025) arises from the large
effective well width value ($\omega/\ell\sim$ 3) in our sample, which
differs somewhat from earlier theoretical works in the literature
where disorder broadening \cite{Morf} or Landau
level mixing\cite{Wojs06} were taken to be the dominant
mechanisms suppressing the experimental gap. We also note that
further improvement (above the $\mu_{A}=28\times10^{6}$cm$^{2}$/Vs
value of sample A) in the sample quality could enhance the gap at
most by a factor of 2 provided mobilities above 50 millions could be
achieved. Our theoretical consideration actually suggests two
alternative (and perhaps simpler) techniques for enhancing the 5/2
gap: (i) Use thinner quantum well samples so that the finite width
correction is smaller; (ii) Use higher carrier density so that the
5/2 FQHE state occurs at higher magnetic field values.

Based on our extensive FQHE activation measurements in the second Landau level, we conclude that (1) the even-denominator 5/2 state, which has no analog in the LLL, is the strongest incompressible state in the SLL; and (2) the 2+ 1/3 (and the related 2+2/3) SLL states are  unlikely to be Laughlin-like states similar to the corresponding 1/3 or 2/3 states in the LLL.  Our measured activation energy for the 7/3 state is an order of magnitude larger than the 12/5 activation energy, but is within a factor of 2 of the 11/5 activation energy. For the LLL Laughlin-like states the situation is precisely reversed with the 1/3 state having an activation energy an order of magnitude (only a factor of 2) larger than the 1/5 (2/5) state. The SLL, in contrast to the LLL where the Laughlin correlation dominates
except at the smallest filling factors, possesses  many competing ground
states of comparable energies for all fillings, considerably complicating
the task of understanding its unique and rich quantum phase diagram.

We thank R.H. Morf for careful reading of the manuscript, W. Pan for useful discussions, and the University of Chicago MRSEC  for  the use of shared facilities. The work is supported by the Microsoft Q Project.



\begin{thebibliography}{99}
\bibitem{Laughlin83} R.B. Laughlin, Phys. Rev. Lett {\bf 50}, 1395 (1983).
\bibitem{Jain89} J.K. Jain,  Phys. Rev. B {\bf 40}, 8079 (1989).
\bibitem{Halperin93} B.I. Halperin, P.A. Lee, and N. Read, Phys. Rev. B {\bf  47}, 7312 (1993).
\bibitem{Pan03} W. Pan, H. L. Stormer, D. C. Tsui, L. N. Pfeiffer, K. W. Baldwin, and K. W. West, Phys. Rev. Lett. {\bf 90}, 16801 (2003).
\bibitem{Willett87} R. Willett, J. P. Eisenstein, H. L. Stšrmer, D. C. Tsui, A. C. Gossard and J. H. English, Phys. Rev. Lett. {\bf 59}, 1776 (1987).
\bibitem{Pan99} W. Pan, J.-S. Xia, V. Shvarts, D. E. Adams, H. L. Stormer, D. C. Tsui, L. N. Pfeiffer, K. W. Baldwin, and K. W. West, Phys. Rev. Lett. {\bf 83}, 3530 (1999).
\bibitem{Xia04} J. S. Xia, W. Pan, C. L. Vicente, E. D. Adams, N. S. Sullivan, H. L. Stormer, D. C. Tsui, L. N. Pfeiffer, K. W. Baldwin, and K. W. West, Phys. Rev. Lett. {\bf 93}, 176809 (2004).
\bibitem{Eisenstein02} J. P. Eisenstein, K. B. Cooper, L. N. Pfeiffer, and K. W. West, Phys. Rev. Lett. {\bf 88}, 076801 (2002).
\bibitem{Moore91} G. Moore and N. Read, Nucl. Phys. B {\bf 360}, 362 (1991).
\bibitem{Greiter91} M. Greiter, X.G. Wen, and F. Wilczek, Phys. Rev. Lett. {\bf 66} 3205 (1991).
\bibitem{Morf98} R.H. Morf, Phys. Rev. Lett. {\bf 80} 1505 (1998).
\bibitem{Rezayi00} E.H. Rezayi and F.D.M. Haldane, Phys. Rev. Lett. {\bf 84}, 4685 (2000).
\bibitem{DasSarma05} S. Das Sarma, M. Freedman and C. Nayak,  Phys. Rev. Lett. {\bf 94}, 166802 (2005); Physics Today {\bf  59}, 32 (2006).
\bibitem{SLLtheory} A.H. MacDonald and S.M. Girvin, Phys. Rev. B {\bf 33}, 4009 (1986); N. d'Ambrumenil and A. M. Reynolds, J. Phys. C. {\bf 21}, 119 (1988); A. Wojs, Phys. Rev. B {\bf 63}, 125312 (2001); F. D. M. Haldane in Quantum Hall Effects, edited by R.E. Prange and S. M. Girvin (Springer, NY, 1985); C. Toke, M.R. Peterson, G.S. Jeon, and J.K. Jain Phys. Rev. B {\bf 72}, 125315 (2005).
\bibitem{Csathy05} G.A. Csathy,  J. S. Xia,  C. L. Vicente, E. D. Adams, N. S. Sullivan, H. L. Stormer, D. C. Tsui, L. N. Pfeiffer, K. W. Baldwin, and K. W. West, Phys. Rev. Lett. {\bf 94}, 146801 (2005).
\bibitem{Du93} R.R. Du, H.L. Stormer, D.C. Tsui, L.N. Pfeiffer, and K.W. West, Phys. Rev. Lett. {\bf 70}, 2944 (1993).
\bibitem{Jiang90} H.W. Jiang, R.L. Willett,  H.L. Stormer, D.C. Tsui, L.N. Pfeiffer, and K.W. West, Phys. Rev. Lett. {\bf 65}, 633 (1990). 
\bibitem{Read99} N. Read and E. Rezayi, Phys. Rev. B {\bf 59}, 8084 (1999).
\bibitem{Morf} R. H. Morf, N. d'Ambrumenil, and S. Das Sarma, Phys. Rev. B {\bf 66}, 075408 (2002); R. Morf and N. d'Ambrumenil, Phys. Rev. B {\bf 68} 113309 (2003).
\bibitem{Feigun} A. E. Feiguin, E. Rezayi, C. Nayak, S. Das Sarma,  arXiv: cond-mat/:0706.4469
\bibitem{Wojs06} A. Wojs and J.J. Quinn. Phys. Rev. B {\bf 74}, 235319 (2006).
\bibitem{dS85} S. Das Sarma and F. Stern, Phys. Rev. B 32, 8442 (1985).

\end{thebibliography}
\end{document}